\documentclass[10pt,letterpaper]{article}
\usepackage{opex3}

\begin{document}

\title{Susceptibility to and correction of azimuthal aberrations in singular light beams}

\author{B R Boruah and M A A Neil}
\address{Blackett Laboratory, Imperial College London, Prince Consort Road, London SW7 2BW}
\email{mark.neil@imperial.ac.uk} 



\begin{abstract} 
We show how the effects of azimuthal optical aberrations on singular light beams can result in an intensity modulation in the beam waist or focal point spread function (PSF) that is directly proportional to the amplitude of the applied phase aberration.  The resulting distortions are enough to significantly degrade their utility even in well corrected optical systems.  However we show that pattern of these intensity modulations is related to the azimuthal order of the applied aberration and we suggest how this can be used to measure those aberrations.  We demonstrate a closed loop system using a liquid crystal spatial light modulator as a programable diffractive optical element to both generate the beam and correct for the sensed aberrations based on feed back from a CCD detected intensity image of the focal point spread function.
\end{abstract}



\section{Introduction}
Singular light beams contain screw phase dislocations across the wave-front often characterised by their topological charge, the multiple of $2\pi$ in phase that the wave traverses along a circular path around the dislocation.  Such beams, sometimes also referred to as optical vortices or orbital angular momentum (OAM) states \cite{allen1992,he1995}, have attracted interest in a wide range of fields including quantum communication cryptography \cite{padgett2002,gibson2004}, optical trapping \cite{curtis2004,fatemi2006} and far field super-resolving optical microscopy \cite{klar2001,hell2006}.  The properties of these beams that make them attractive for such applications include the sharp zero of intensity that occurs at the dislocation, the circularly symmetric donut intensity distribution they result in when focussed and the fact that modes with different topological charge can be considered as distinct quantum states of light.  Unfortunately, when optical aberrations are superimposed on these beams, for instance by passing through the atmosphere or simply due to optical misalignments, these properties can be severely degraded.  For instance, it has been shown \cite{paterson2005} that aberrations due to atmospheric turbulence results in severe mixing of OAM states, sufficient to destroy the information in a quantum encoded data stream.  

In this paper we concentrate on applications where the singular beam is focused with a lens.  In the ideal case the resulting focal point spread function consists of a bright ring of intensity surrounding a sharp zero of intensity.  However, we show that this ideal case is quickly degraded even in what is normally considered to be well corrected optics.  Indeed it is rare to see experimental results presenting focussed singular beams that are not imperfect or distorted in some way. We show analytically that the focal point spread functions of singular beams have  off-axis linear sensitivity to both odd and even azimuthal aberrations while those of non-singular beams have off-axis linear sensitivity only to odd azimuthal aberrations. We utilize this property of the singular beam to detect the corresponding aberrations using a segmented detector. We then implement this technique in a closed-loop system based on feedback from a CCD camera acting as the detector, using a liquid crystal spatial light modulator as the beam shaping device.

\section{The sensitivity of singular and normal beams to azimuthal aberrations}

\begin{figure} [htb]
\centering\includegraphics[width=10cm]{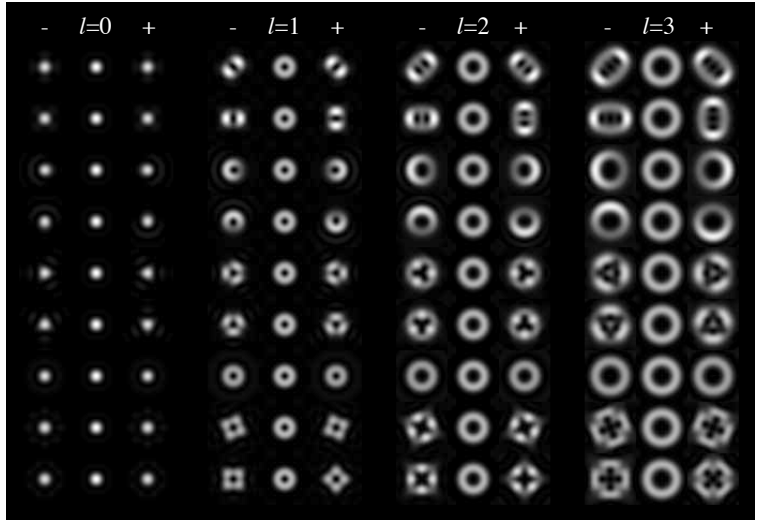}
\caption{\label{fig_1} Intensity patterns in the Fourier plane for a normal beam ($l=0$) and singular beams ($l= 1, 2, 3$) aberrated with  $-0.447Z_i$ (left column) and $+0.447Z_i$ (right column) of $Z_{i=5, 6, 7, 8, 9, 10, 14, and\ 15}$ (from top to bottom).}
\end{figure}

Fig.\ref{fig_1} illustrates the effect of aberrations on the point spread function of a simple lens with its circular pupil filled with a constant intensity singular beam of varying topological charge, $l$.  The intensity point spread function is evaluated in the paraxial scalar limit by calculating the modulus squared of the Fourier transform of the pupil function for various applied phase aberrations corresponding to the lowest order Zernike aberration modes.  Zernike polynomials $Z_i$ are often used to describe optical aberrations over a circular aperture, \cite{mahajan}, and can be expressed as a the product of a radially varying component $R_i(r)$ and an azimuthally varying component $T_i(\theta)$.  In fig.\ref{fig_1} an amount of aberration for each Zernike mode has been applied to result in a root mean square phase aberration of $\phi_{RMS}=0.447$ radians, sufficient to reduce the Strehl ratio of an otherwise perfect lens to 0.8 - the minimum figure considered for a well corrected system\cite{born}.  The resulting PSFs for a planar beam are also shown and clearly exhibit limited distortions compared to those visible on the singular beams.  For instance astigmatism distorts the donut intensity pattern into a two lobed pattern.  Indeed, astigmatism can be used to convert between Hermite Gaussian modes and singular Laguerre Gaussian modes using cylindrical optics \cite{tamm1990, mw1993}.

The sensitivity of the intensity distribution in a PSF to aberrations can be considered in terms of how that intensity varies as the amplitude of the applied aberration is increased.  For instance, an unaberrated planar beam produces an absolute peak of intensity on-axis that can only decrease as any aberration is applied.    To maintain an analytic form this means that the intensity varies only quadratically with the amplitude of the applied aberration.  Conversely, looking at points off-axis, the intensity is not constrained at a maximum value and so can vary linearly with the amplitude of the applied aberration.  It is this linear sensitivity that results in significant distortion even at low levels of aberration.  Of course the off-axis intensity of a planar beam is reduced relative to that on-axis and so any such distortions are not so significant in terms of the whole PSF.  However, for singular beams the intensity in the PSF  is concentrated off-axis and these distortions can become highly significant.    We analyse these issues by calculating the sensitivity, $s_i$, of a singular or normal beam to Zernike aberration mode, $Z_i$, as the linear slope of the PSF intensity, $I$, with applied aberration amplitude, $a_i$.

The amplitude PSF, $U(\rho,\xi)$, is given by the Fourier transform of the unit radius pupil function $P(r,\theta)$:
\begin{equation}
\label{eq_1}
  U(\rho,\xi)=\mathcal{F}\left[P(r,\theta)\right]=\mathcal{F}\left[e^{jl\theta}e^{ja_iZ_i(r,\theta)}\right]
\end{equation} 
where $\mathcal{F}\left[\right]$ is the Fourier transform over the unit circle from cylindrical co-ordinate space $\left(r,\theta\right)$ to cylindrical co-ordinate space $\left(\rho,\xi\right)$ given as
\begin{equation}
\label{eq_2}
\mathcal{F} [f(r,\theta)]=\int_0^{2\pi}\!\!\!\!\!\int_0^1f(r,\theta)e^{j\rho r cos(\theta-\xi)}r \,dr d\theta\\
\end{equation}
The PSF intensity is then written as   
\begin{equation}
\label{eq_3}
I(\rho,\xi)=U(\rho,\xi)U^*(\rho,\xi)
\end{equation}
Hence according to our definition the sensitivity $s_i$ can be written as  
\begin{equation}
\label{eq_4}
  s_i  =  \left.  \frac{dI}{da_i}\right|_{a_i=0} = 2\Re{\left[ U(\rho,\xi)  \frac{dU^*(\rho,\xi)}{da_i}\right]}\bigg|_{a_i=0}
\end{equation} 
Substituting equation eq.\ref{eq_1} into eq.\ref{eq_4} one obtains:
\begin{equation}
\label{eq_5}
  s_i  =  2\Re\left[ \mathcal{F}\left[e^{jl\theta}\right]\left(\mathcal{F}\left[jZ_i(r,\theta)e^{jl\theta}\right]\right)^*\right]
\end{equation} 
Zernike mode polynomials have the general form $Z_i(r,\theta)=R_n(r) \cos(m\theta)$ or $Z_i(r,\theta)=R_n(r) \sin(m\theta)$ with radial variation $R_n(r)$ and integer value $m$ depending on the mode index, $i$.  Note that circularly symmetric modes such as defocus and spherical aberration are represented by modes of the first form with $m=0$.  Using the Fourier transform properties of such functions in polar co-ordinates \cite{neil2000}, we can write
\begin{equation}
\label{eq_6}
\mathcal{F} [R_n(r)e^{j m \theta}]=2\pi j^m e^{j m \xi } \int_0^1{R_n(r)J_m(\rho r)r dr}\\
\end{equation}
where $J_m$ is the Bessel function of the first kind of order $m$.  Eq.\ref{eq_5} can be simplified using eq.\ref{eq_6}, and basic trigonometry to give expressions for $s_i$ as follows.  
\\
\\With $Z_i(r,\theta)=R_n(r) \cos(m\theta)$
\begin{eqnarray}
 \label{eq_7}
s_i&=&(-1)^{\frac{m}{2}}  F_{l} \left( (-1)^{l} G_{m-l} - G_{m+l} \right) \sin(m\xi) \quad \mbox{$m$ even} \\
 \label{eq_8}
s_i&=&(-1)^{\frac{m+1}{2}} F_{l} \left( (-1)^{l} G_{m-l} + G_{m+l} \right) \cos(m\xi) \quad \mbox{$m$ odd} 
\end{eqnarray}
And similarly with $Z_i(r,\theta)=R_n(r) \sin(m\theta)$
\begin{eqnarray}
 \label{eq_9}
s_i&=&(-1)^{\frac{m}{2}}  F_{l} \left( -(-1)^{l} G_{m-l}+ G_{m+l} \right) \cos(m\xi)  \quad \mbox{$m$ even} \\
 \label{eq_10}
s_i&=&(-1)^{\frac{m+1}{2}}    F_{l} \left( (-1)^{l} G_{m-l}+ G_{m+l} \right) \sin(m\xi) \quad \mbox{$m$ odd} 
\end{eqnarray}
where 
\begin{eqnarray}
\label{eq_11}
F_{k}&=&2\pi \int_0^1{J_k(\rho r)r dr} \\
\label{eq_12}
G_{k}&=&\pi\int_0^1{R_n(r) J_k(\rho r)r dr} 
\end{eqnarray}

Substituting $m=0$ in eq.\ref{eq_7}, it can be seen that $s_i$ vanishes for all values of $l$, hence all modes are insensitive to spherical aberrations.  For $l=0$ beams (no singularity) it can be seen that equations eq.\ref{eq_7} and eq.\ref{eq_9} yield $s_i=0$ for even \textit{m}.  For odd \textit{m} and $l=0$ beams, the sensitivity on-axis ($\rho=0$) where the intensity in the PSF is highest, must also be zero from eq.\ref{eq_12}.  Hence a non-singular beam is only sensitive to odd aberrations, and then only off-axis where the intensity is low.  Conversely for singular beams ($l\ne0$) and non-spherical aberrations ($m\ne0$), $s_i$ always shows a $\cos(m\xi)$ or $\sin(m\xi)$ functional form off-axis (and in some cases on-axis) where the power in the singular beam is concentrated.  It is this result that demonstrates the linear sensitivity of PSF intensity to azimuthal aberrations for singular beams.

\section{Design of the sensor detector } 

Discussion in the above section suggests that by analysing the azimuthal harmonic content of the PSF it should be possible to determine the azimuthal content of the aberrations present on the beam, which could then be used to provide a feedback signal for correction of those aberrations.  Rather than using a whole image transform approach to achieve this, we suggest a segmented detector approach that would be more applicable to simple or low light level applications.  This form of detector would be easy to implement as a simple segmented photodiode, and with relatively few steps required for computation of the aberration signals, would be applicable to real-time aberration correction applications.
\begin{figure} [htbp]
\centering\includegraphics[width=12cm]{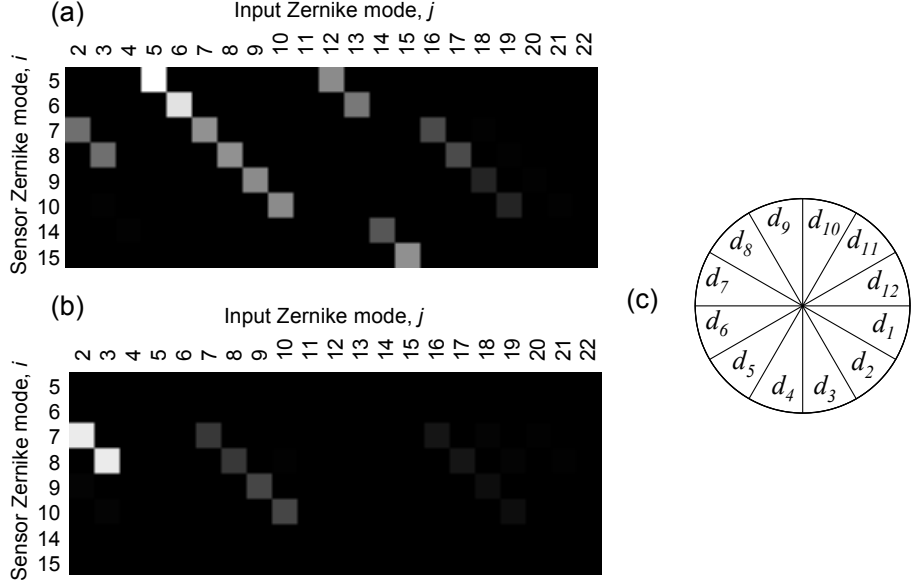}
\caption{\label{fig_2} Gray scale images representing the magnitude of the sensitivity matrix, $\left|S_{ij}\right|$ for (a) as singular $l=1$ beam and (b) a non-singular $l=0$ beam.  The displayed gray scale is normalised such that the maximum value, $S_{55}$ with $l=1$, is set to white.  (c) The segmented sensor detector pattern.}
\end{figure}

The sensor comprises twelve detectors, $d_n$, each equal to a segment of angular width $\frac{\pi}{6}$ radian of a reference circle as shown in fig.\ref{fig_2}(c) and generating a signal $\delta_n$ ($n= 1 \rightarrow 12$). The reference circle and its radius are termed as the sensor circle and sensor radius respectively. The sensor circle is symmetrically positioned in the focal plane of the aberrated singular beam. The detector signals from the sensor are split into two groups for each Zernike mode, namely a positive group, $n^+$ and a negative group, $n^-$, as given in table \ref{table_1}. 
\begin{table}
\begin{center}
\begin{tabular}{|c|c|c|c|c|c|}
\hline
$Z_i$&$n^+$&$n^-$&$S_{ii} (l=1)$&$S_{ii} (l=0)$&$\sigma_{ii}\times\sqrt{N}$\\
\hline
$Z_5$&4,5,6,10,11,12&1,2,3,7,8,9&-1.0&0&0.93\\
$Z_6$&1,6,7,12&3,4,9,10&-0.89&0&0.86\\
$Z_7$&5,6,7,8&1,2,11,12&-0.51&-0.18&1.5\\
$Z_8$&8,9,10,11&2,3,4,5&-0.51&-0.18&1.5\\
$Z_9$&2,3,6,7,10,11&1,4,5,8,9,12&-0.49&-0.22&2.0\\
$Z_{10}$&3,4,7,8,11,12&1,2,5,6,9,10&-0.49&-0.22&2.0\\
$Z_{14}$&1,4,7,10&3,6,9,12&-0.28&0&2.7\\
$Z_{15}$&2$\times$(2,5,8,11)&1,3,4,6,7,9&\ -0.51\ &0&2.6\\
\hline
\end{tabular}
\caption{Geometry and performance parameters for the proposed wave-front sensor}\label{table_1}
\end{center}
\end{table}
\begin{equation}
\Delta_{ij}=\frac{\sum \delta_{n^+} - \sum \delta_{n^-}}{\sum \delta_n}
\label{eq_13}
\end{equation}
\begin{equation}
S_{ij}=\left. \frac{d{\Delta_{ij}}}{da_j}\right|_{a_j=0}
\label{eq_14}
\end{equation}
The sensor signal, $\Delta_{ij}$, is defined as the response of the $i$th sensor, designed to detect  Zernike mode $Z_i$, to the presence of Zernike mode $Z_j$. This is obtained by subtracting the total number of photons collected in the negative half from that collected in the positive half, and then normalizing by the total number of photons collected, as shown in eq.\ref{eq_13}, (note that for $Z_{15}$ the $n^+$ signals are counted twice). With this the sensor is able to detect 8 non-rotationally symmetric Zernike modes ($Z_{i=5, 6, 7, 8, 9, 10, 14, and\ 15}$), referred to henceforth  as the 8 sensor modes.  We also define the sensitivity $S_{ij}$ of the $i$th sensor signal to the $j$th mode, $Z_j$, as the slope of the sensor signal vs strength of aberration, $a_j$, at $a_j=0$, as shown in eq.\ref{eq_14}.

\section{Numerical simulation of the sensor performance}

\begin{figure} [htbp]
\centering\includegraphics[width=12cm]{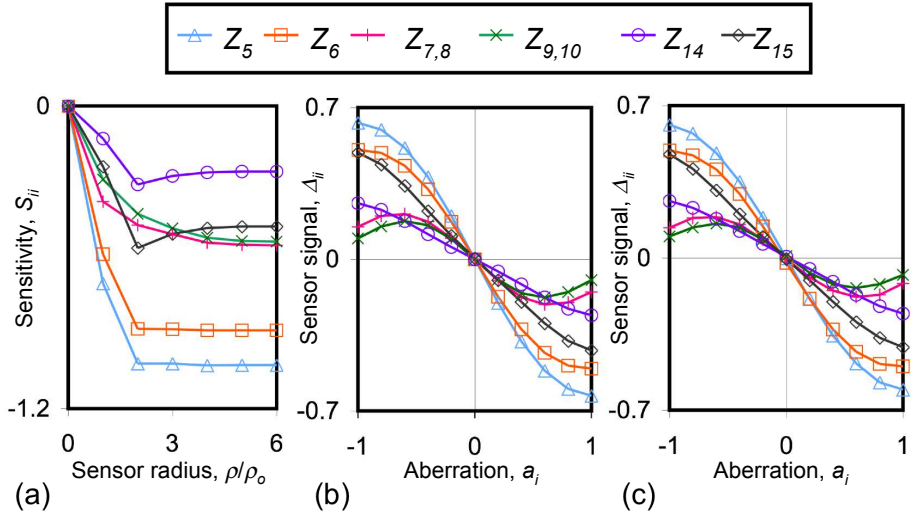}
\caption{\label{fig_3} (a) Sensitivity, $\Delta_{ii}$, as a function of normalised sensor radius, $\rho/\rho_o$. (b) Simulated  and (c) experimental sensor signal, $\Delta_{ii}$, vs aberration, $a_i$, for $Z_{5, 6, 7, 8, 9, 10, 14, 15}$.}
\end{figure}

The performance of the sensor was simulated numerically using Fourier techniques according to eq.\ref{eq_1} to determine an appropriate sensor circle radius.  The sensitivity of the sensor for each sensor mode was computed as a function of sensor radius, normalised by the radius of the first zero of the Airy disk, $\rho_o$.  The resulting self sensitivities, $S_{ii}$, are plotted as a function of the normalised sensor radius, $\rho/\rho_o$, in fig.\ref{fig_3}(a).   It can be seen that for different  sensor modes the sensitivity saturates at a radius  $\rho=3\rho_o$.  The self sensitivity, $S_{ii}$, at this radius is given in table \ref{table_1} and we use this radius for all our subsequent calculations and experiments.  The full sensitivity matrices were then calculated for the sensor against all Zernike modes up to $j=22$ (second order spherical aberration) with both singular ($l=1$) and normal ($l=0$) beams shown graphically in fig.\ref{fig_2}(a) and fig. \ref{fig_2}(b) respectively.  The self sensitivities with singular beam are strong with a certain amount of cross sensitivity present, not least due to our chosen simplified method of detection and sensor signal calculation. As suggested by eq.\ref{eq_7} and eq.\ref{eq_9}, the self sensitivities with normal beam vanish for sensor modes with an even value of $m$, and are typically significantly lower than those for the singular beam. 

The sensor signal was also calculated for each sensor mode over the range $-1<a_i<1$ and is shown in fig.\ref{fig_3}(b).  The sensor gives a variable performance depending on the particular mode being sensed, the best being for $Z_5$ and $Z_6$ (astigmatism).  It is also noted that the effective linear range of the sensor is relatively short with saturation appearing between $0.5<\left|a_i\right|<1.0$.  The range of the sensor is limited, but this is a result of and should be considered along side the high sensitivity produced.  This high sensitivity, although over a short range, is particularly useful when used in closed loop control systems.

Along with the sensor signal the associated error in the sensed aberration is estimated analytically based on the shot noise in the sensed photon stream, assuming a mean number of $N$ photons and a Gaussian approximation to the Poissonian noise statistics.  This estimated error, $\sigma_{ii}$, is also shown in table \ref{table_1}.  

\section{Experimental implementation}
\begin{figure} [htbp]
\centering\includegraphics[width=8.5cm]{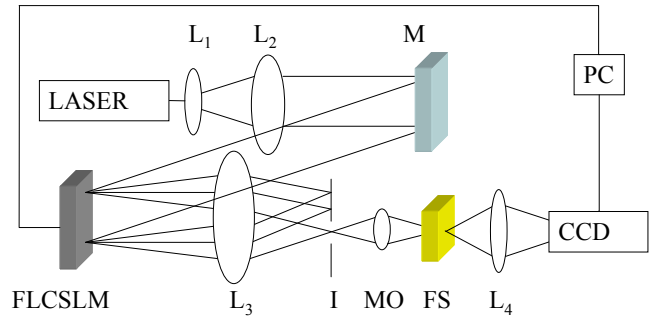}
\caption{\label{fig_4} Experimental set up for closed-loop aberration correction.}
\end{figure}

A proof of principle system was constructed as shown in fig.\ref{fig_4}.  Beam expanded and collimated light from a laser system operating at 532nm illuminates a reflective ferro-electric liquid crystal spatial light modulator (FLCSLM - CRL Opto SXGA).  A computer generated hologram, designed using the technique described in \cite{neil1998, neil2000m}, is displayed on the device over a circular area of diameter 512 pixels.  The hologram can be dynamically updated from a LABVIEW program running on a PC and, along with lens L3 and iris I, effectively allows the accurate generation of arbitrarily shaped wavefronts across the 512 pixel system pupil.  Light reflected from the FLCSLM is directed towards lens L3 where the PSF of the system is realised in its back focal plane on the +1 diffracted order of the computer generated hologram.  This beam is then imaged on to a fluorescence slide (FS), emitting in the red, with a 10X microscope objective (MO).  A lens $L_4$ forms an incoherent image of the resulting intensity PSF on the CCD camera (Hammamatsu ORCA ER) panel.  A red filter (F) in front of the camera ensures that it receives only fluorescent light from the slide.  This system allows the capture of clean images without coherent artefacts caused by scatter or reflections in the CCD camera.  Fig.\ref{fig_4}(a) shows a comparison between the experimentally generated and simulated PSFs for a singular ($l=+1$) mode with $0.447Z_i$ of various Zernike aberrations applied, clearly demonstrating the effectiveness of the system. 

\begin{figure} [htbp]
\centering\includegraphics[width=10cm]{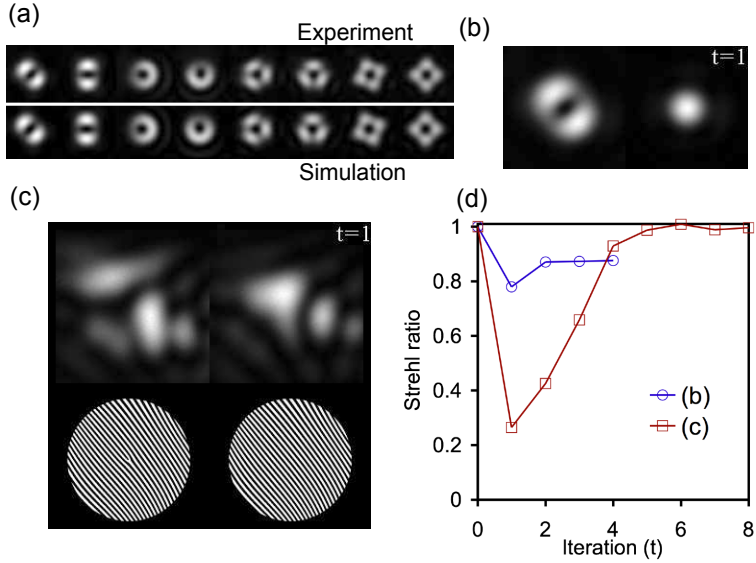}

\caption{\label{fig_5} (a)Experimental (top) and simulated (bottom) intensity PSFs for the $l=1$ singular beam aberrated with 0.447X Z$_{i=5, 6, 7, 8, 9, 10, 14, 15}$ (left to right). (b) (92 KB) $l=1$ PSFs (left) and corresponding normal $l=0$ PSFs (right) as a sequence (t=0$\rightarrow$4) during the correction procedure for a well corrected beam (t=0), aberrated by a glass slide. (c) (320 KB) Sequence (t=0$\rightarrow$8) of $l=1$ (top left), $l=0$ (top right) PSFs and the corresponding holograms (bottom) during the correction of a heavily aberrated beam. (d) Peak intensity of the $l=0$ PSFs as a Strehl ratio vs iteration, for the images shown in  (b) and (c).}
\end{figure}

In the system as realised the radius of the Airy disk was approximately $\rho_o=20$ pixels and so a sensor radius of 60 pixels was chosen.  Captured images were read by a LABVIEW program running in the PC and the sensor signal was calculated by the program according to eq.\ref{eq_13}.  A small error is introduced in the sensed aberrations when matching the detector segments to the pixellated CCD array.  With 11,096 pixels in the sensor circle we estimate this error to be less than 0.0022 radians for each of the sensor modes.  Alignment of the detector with respect to the centre of the unaberrated beam is also a crucial requirement for the detection of correct signals.  The lateral offset of the PSF can be considered in terms of the particular Zernike modes $Z_2$ and $Z_3$ which represent tilts of an otherwise planar wavefront in the pupil plane of the lens.  However, when using an $l=0$ beam fig.\ref{fig_2}(b) shows that the sensor signals  $\Delta_7$ and $\Delta_8$ become predominantly sensitive to  $Z_2$ and $Z_3$ ($S_{72}=S_{83}=-0.92$) and can therefore be used to centre the beam on the sensor circle.

The performance of the sensor was evaluated experimentally by determining the sensor signal for each sensor mode as the amplitude of that mode was varied.  The results are plotted in fig.\ref{fig_3}(c) showing clear agreement with the simulation results in fig.\ref{fig_3}(b).

Closed-loop control of aberrations was also demonstrated by using the detected sensor signals as the feedback correction factor in a simple proportional closed loop control system.  The movie in fig.\ref{fig_5}(b) shows an initially well corrected beam at iteration $t=0$ that is aberrated at $t=1$ by placing a glass microscope slide in the beam path. For this small, mainly astigmatic aberration the correction is made in a single step and in the following 3 iterations the PSF remains stable with only a residual drop in Strehl attributed to reflections from the glass slide. This Strehl ratio is plotted at time $t=0$ and subsequent times in fig.\ref{fig_5}(d).

As a second demonstration of closed loop operation a random mix of the sensor modes, totalling $\phi_{RMS}=1.4$ radians, was applied to the system as an internal initial offset.  The subsequent evolution of the system PSF along with the corresponding holograms is shown in the movie in fig.\ref{fig_5}(c), and the Strehl ratio is plotted in fig.\ref{fig_5}(d).  Again the correction of the applied aberrations is clearly demonstrated in about 4 iterations, with the system exhibiting stability over the remaining 3 iterations.  It is noted that in both these cases we considered only the diagonal elements of the sensitivity matrix $S_{ij}$ and made no effort to optimise the control loop for speed or stability.

\section{Conclusion}
We have shown that singular beams are ultra sensitive to non-rotationally symmetric aberrations, such that even well corrected  optics produces significant distortion of the beam.  We also show that this distortion manifests itself as an intensity modulation in the PSF of the system with the same azimuthal dependance as an applied Zernike aberration mode.  We have suggested how this can form the basis of an aberration sensor that can measure those aberrations and experimental results demonstrate the efficacy of our aberration sensor to correct non-rotationally symmetric aberrations. The aberration sensor also shows advantages in comparison to other sensing techniques in that there may be no need to include extra components in the system than are already there.  Indeed it could even be considered to provide enough sufficient visual feedback for the manual correction of aberrations that have been introduced by simple component misalignment. 
\section*{Acknowledgments}
The authors wish to acknowledge the financial support from the EPSRC, UK and the Paul Foundation, India.

\end{document}